\documentclass{webofc}
\usepackage[varg]{txfonts}   
\begin{document}
\title{Multi-scale dynamics in star-forming regions: the interplay between gravity and turbulence}
%
%

\author{\firstname{Alessio} \lastname{Traficante}\inst{1}\fnsep\thanks{\email{alessio.traficante@inaf.it}} \and
        \firstname{Gary A.} \lastname{Fuller}\inst{2} \and
        \firstname{Ana} \lastname{Duarte-Cabral}\inst{3} \and
        \firstname{Davide} \lastname{Elia}\inst{1} \and
        \firstname{Mark H.} \lastname{Heyer}\inst{4} \and
        \firstname{Sergio} \lastname{Molinari}\inst{1} \and
        \firstname{Nicolas} \lastname{Peretto}\inst{3} \and
        \firstname{Eugenio} \lastname{Schisano}\inst{1}
        }

\institute{IAPS - INAF, via Fosso del Cavaliere, 100, I-00133 Roma, Italy
\and
Jodrell Bank Centre for Astrophysics, Department of Physics and Astronomy, The University of Manchester, Oxford Road, Manchester M13 9PL, UK
\and
School of Physics and Astronomy, Cardiff University, Queens Buildings, The Parade, Cardiff CF24 3AA, UK
\and
Department of Astronomy, University of Massachusetts, Amherst, MA 01003, USA
          }

\abstract{%
In the multi-scale view of the star formation process the material flows from large molecular clouds down to clumps and cores. In this paradigm it is still unclear if it is gravity or turbulence that drives the observed supersonic non-thermal motions during the collapse, in particular in high-mass regions, and at which scales gravity becomes eventually dominant over the turbulence of the interstellar medium. To investigate this problem we have combined the dynamics of a sample of 70 $\mu$m-quiet clumps, selected to cover a wide range of masses and surface densities, with the dynamics of the parent filaments in which they are embedded. We observe a continuous interplay between turbulence and gravity, where the former creates structures at all scales and the latter takes the lead when a critical value of the surface density is reached, $\Sigma_{th}= 0.1 $ g cm$^{-2}$. In the densest filaments this transition can occur at the parsec, or even larger scales, leading to a global collapse of the whole region and most likely to the formation of the massive objects.
}
\maketitle
\section{Introduction}
\label{intro}

The formation of stars is a multi-scale process which starts in giant molecular clouds, structures extended tens of parsecs. Within these clouds the process continues through dense, elongated structures called filaments and, within these filaments, parsec-scale dense clumps that are the progenitors of sub-parsec cores and fragments, finally leading to the formation of (massive) cluster of stars. In this scenario it is still unclear how the interplay between gravity and turbulence regulates this continuous flow of energy and gas across several orders of magnitude of spatial scales and gas densities. The theoretical scenarios are still divided in models where the turbulent flow is primarily responsible for the formation of the parsec and sub-parsec structures \cite{Padoan}, and opposite approaches in which gravity takes over the turbulent cascade already at the scales of several parsecs \cite{Vazquez-Semadeni}. 

Observationally, the study of such flows consider two critical aspects: 1) the observations of star forming regions with various physical properties, such as mass and densities, to be able to probe the interplay of turbulence and gravity under different gravitational potentials; 2) the observations of tracers sensitive to the progressively denser layers of each star-forming region, which allow us to observe the gas dynamics from the (relatively) diffuse clouds observed with low-J transitions of CO to the dense, massive clumps observed with high-density tracers such as N$_{2}$H$^{+}$ (1-0). Several works have been dedicated to the study of the gas dynamics in statistically significant sample of objects traced with a specific tracer, such as the GMCs observed with CO (1-0) \citep{Miville-Deschenes}, filaments with e.g. CO (2-1) \cite{Duarte-Cabral} or CO (3-2) \cite{Rigby} and massive clumps with N$_{2}$H$^{+}$ (1-0) \cite{Traficante}. In this work we present a different approach: we show the results obtained from the combination of different gas tracers used to look within the same star-forming regions with the aim of investigating the turbulent cascade and the role of the gravitationally driven motions at the different density layers in each region.

\section{Dataset}
\label{sec:data}
In order to study the turbulent cascade and its interplay with the gravitational potential, it is mandatory to minimize any possible processes that can alter the gas kinematics. The starting requirement to perform our study has been the choice of a sample of 70 $\mu$m-quiet clumps, i.e. objects at the earliest phases of their evolution and therefore likely to be unaffected by internal feedback from newborn stars \cite{Traficante20}. The lack of the 70 $\mu$m emission indeed guarantees very low value of the luminosity over mass (L/M) parameter in these objects, which is a strong indicator that the clump is almost pristine, with very little or no internal star formation yet started \cite{Molinari}. Our final sample consists of 22 clumps which have been further chosen to be divided into three different surface density bins: $\Sigma_{low}$, i.e. clumps with surface density $\Sigma<0.05$ g cm$^{-2}$ (5 clumps); $\Sigma_{int}$, i.e. clumps with $0.05\leq\Sigma\leq 0.1$ g cm$^{-2}$ (6 clumps); $\Sigma_{high}$, i.e. clumps with $\Sigma> 0.1$ g cm$^{-2}$ (11 clumps). These values are all above the suggested threshold for star-formation ($\simeq0.02$ g cm$^{-2}$, \cite{Lada}) and the subdivision follows the suggested thresholds for massive star-formation ($\Sigma\geq 0.05$ g cm$^{-2}$, \cite{Urquhart}; $\Sigma\geq 0.1 $ g cm$^{-2}$, \cite{Tan}). This sample therefore allows us to explore the interplay between gravity and turbulence in different potential wells. The kinematics of these clumps have been evaluated with N$_{2}$H$^{+}$ (1-0) observations collected by the IRAM 30m single dish telescope. In addition, for each clump we identify the parent filament in the Hi-GAL filaments catalogue \cite{Schisano}, from which we extracted the mass and equivalent radius of each object. The kinematics of the filaments are evaluated using the $^{13}$CO (1-0) emission. We carefully mask the positions and solid angles of all clumps where the CO emission is likely optically thick in order to measure the average velocity dispersion of the filament itself as opposed to the kinematics of the inner clumps evaluated using the N$_{2}$H$^{+}$ (1-0) emission. Furthermore, we estimate and subtract the contribution of the large-scale velocity gradients from each filament to derive the velocity dispersion from turbulence and/or gravitational motions.

\section{Results}
\label{sec:results}
The first result to notice is that the high density clumps are embedded in the densest filaments of our sample \cite{Traficante20}. In addition, the internal velocity dispersion of the $\Sigma_{int}$ and $\Sigma_{high}$ clumps correlates very well with the velocity dispersion of the parent filament after the gradient subtraction \cite[Pearson correlation coefficient $\rho=0.79$]{Traficante20}. These results suggest that the environment plays a significant role in the formation of the massive objects and the dynamics of the clumps are inherited from the dynamics of their parent cloud. The different scales implies a continuous flow of material from the cloud down to the star-formation sites. 

\begin{figure}[t]
\centering
\sidecaption
\includegraphics[width=7cm,clip]{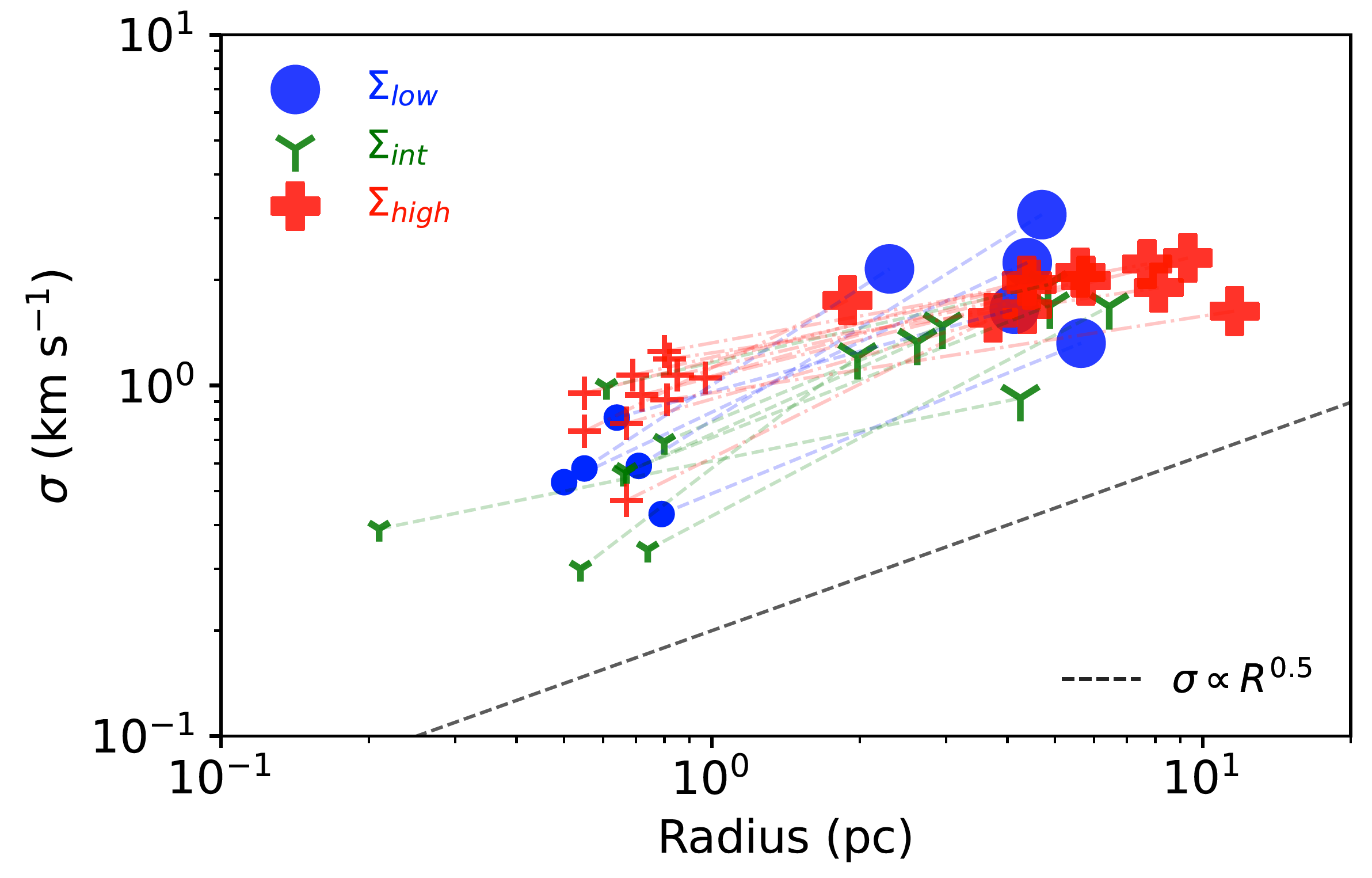}
\caption{Larson relation for the clumps and the parent filaments belonging to the three groups. The dotted lines connect each clump (left-hand points) with the parent filament (right-hand points). Objects associated with clumps in the $\Sigma_{low}$, $\Sigma_{int}$ and $\Sigma_{high}$ groups are represented as blue circles, green tri\_downs and red crosses, respectively. The black-dotted line is the $\sigma\propto$ R$^{0.5}$ relation in the case of a supersonic turbulent cascade of energy from large down to small scales \cite{Traficante20}.}
\label{fig:Larson_relation}       
\end{figure}

In order to investigate the interplay between gravity and turbulence at the different scales, we built the $\sigma$ vs. radius (R) relation for both filaments and clumps separately \cite{Traficante20}. These plot do not show the expected relation for supersonic turbulence, i.e.  $\sigma\propto R^{\delta}$ with $\delta=0.5$ \cite{Larson}. The break of this relation was already noticed at the clump scales in previous surveys of massive clumps \cite{Ballesteros-Paredes, Traficante}, and it has been interpreted as evidence of gravity dominating over turbulence at these scales. With this work we point out that the role of environment can actually hide the turbulent cascade in some cases, in particular in regions where the local environment is naturally less turbulent than others. To demonstrate this conjecture we show the $\sigma$ vs. R plot but this time connecting the clump position in this plot with the position of its parent filament (see Figure \ref{fig:Larson_relation}) . As discussed more in details in \cite{Traficante20}, the average values of $\delta$ of the lines connecting $\Sigma_{low}$, $\Sigma_{int}$ and $\Sigma_{high}$ clumps to their parent filaments are, respectively, $\delta_{low}=0.68$, $\delta_{int}=0.60$ and $\delta_{high}=0.38$. These numbers indicate that when the turbulent cascade is investigated within the same object, and therefore within the same local conditions of the environment, the effect of a turbulent cascade may be identified. The average increasing of the exponent in denser and denser regions is interpreted as evidence that in clumps with $\Sigma> 0.1 $ g cm$^{-2}$ there must be an extra contribution to the kinetic energy, likely driven by the gravitational collapse, which dominates over the energy that could arise from a pure turbulent cascade \cite{Traficante20}. The critical threshold above which gravity overtakes the turbulent cascade may br driven by the local density of the medium, rather than by a specific spatial scale. 

\section{Discussion and conclusions}
\label{sec:conclusion}
\begin{figure}
\centering
\sidecaption
\includegraphics[width=5cm,clip]{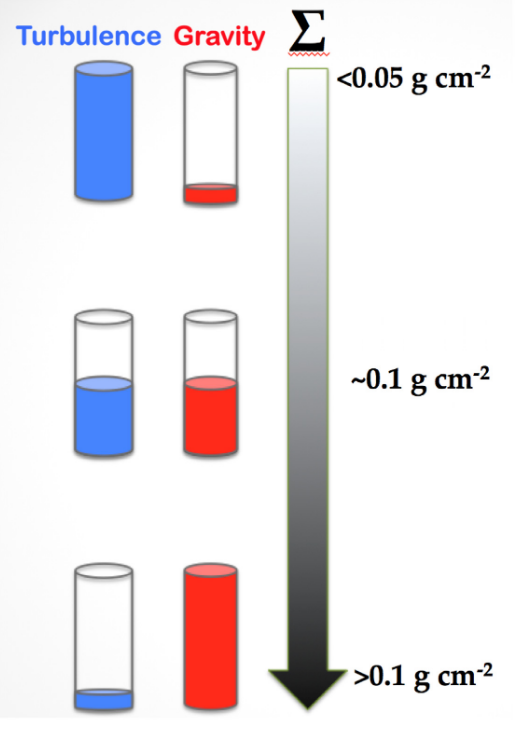}
\caption{Cartoon representation of the interplay between gravity and turbulence in star-forming regions. A cloud begins to be compressed due to the effect of the strong interstellar medium turbulence, with a minimal contribution from gravity. As the material starts to accumulate in different eddies due to turbulence, these regions increase their surface density and with it the role of gravity in driving the motions. As the density reaches the critical value of $\Sigma_{th}= 0.1 $ g cm$^{-2}$, the motions induced by gravity dominate over the turbulence, which, from this point, becomes sub-dominant \cite{Traficante20}.}
\label{fig:model_scenario}       
\end{figure}

This work concludes that the dynamics of the filaments and of their embedded clumps are strongly correlated. The flow of energy and gas is continuous from the filament down to the clumps, and the interplay between gravity and turbulence depends on the density of the objects under investigation. We propose a scenario where the transfer of energy via the turbulent cascade is the dominant mechanism on the largest, and less dense, scales, with a gradual variation in the dynamics: the gravity gains more and more importance as we move into denser and denser regions (e.g. \cite{Pingel}). Above the critical threshold of $\Sigma_{th}= 0.1 $ g cm$^{-2}$ gravity dominates the motions and the kinematics over the turbulent cascade, driving the global collapse of the region above this density threshold. This critical value of the surface density can be reached at any spatial scales, depending on the initial conditions of the turbulent cloud. Gravity can therefore start to drive the motions at the core, the clump, or even the filament scales in star-forming regions embedded in significantly dense clouds. The cartoon in Figure \ref{fig:model_scenario} describes the scenario that we propose, which is detailed in \cite{Traficante20}.




%
%
%

\end{document}